\documentclass[twocolumn,twoside,slac_two]{revtex4}
\usepackage{graphicx}
\usepackage{fancyhdr}
\begin{document}

\title{Interference Effects in $D$ Meson Decays}

%

\author{D. Cinabro}
\affiliation{Wayne State University, Detroit, MI, USA}

\begin{abstract}
I review recent results on Dalitz plot analyses of $D$ meson decays
and an analysis that takes advantage of the quantum correlations between
$D$ meson pairs produced in the decay of the $\psi(3770)$.
\end{abstract}

\maketitle

\thispagestyle{fancy}


\section{Introduction}

	Taking advantage of quantum effects in meson decays provides
a unique window on the mechanics of such decays.  The Dalitz
plot analysis technique~\cite{Dalitz} is a unique way to measure
decay amplitudes and phases.  Not only are these valuable
inputs and a challenge to QCD, but also these parameters
are necessary to extract fundamental parameters.  Specifically
see Jean-Pierre Lees' contribution to these proceedings
on the measurement of the CKM angle $\gamma(\phi_3)$ in 
charged $B$ decays using a Dalitz plot analysis of
$D^0 \to K_{\rm S}\pi^+\pi^-$.

	In threshold production via resonance decays, the
favored method for $e^+e^-$ meson factories, the product mesons
are in an eigenstate of~C.  Specifically the decay of the $\psi(3770)$
produced in $e^+e^-$ collisions are~C~$-$.  This introduces
correlations between the decays of the $D$ mesons produced in the
$\psi(3770)$ decay.  These depend on amplitudes for doubly Cabibbo
suppressed(DCS) decays, $D$-mixing parameters, and strong phases all
of which can be measured in samples of $\psi(3770)$ decays in which
both $D$ decays have been identified as a flavor tag, a semileptonic
decay, or a CP eigenstate~\cite{AsnerandSun}.

	I review results presented since the 2005 summer conference
season.  There are not a large number with all of them being preliminary
results from the CLEO collaboration.  The various versions
of the experiment are described in detail elsewhere~\cite{CLEOdet}.
The work discussed here is based on 9/fb of $e^+e^-$ data collected by
the CLEO-III detector operating in the region of the $Upsilon(4S)$,
and 281/pb of $e^+e^-$data collected by the CLEO-c detector operating
at the $\psi(3770)$, corresponding to 1.4 million $D \bar D$ pairs.

\section{Dalitz Plot Analysis of $D^0 \to K^+K^-\pi^0$}

	The motivation for this analysis is to measure the strong phase
in $D \to K^\ast K$ decays.  A constraint on this phase can be used in the
measurement of the CKM angle $\gamma(\phi_3)$ in charged $B$ decays
as discussed here~\cite{gamma}.

\begin{figure*}[t]
\centering
\includegraphics[width=135mm]{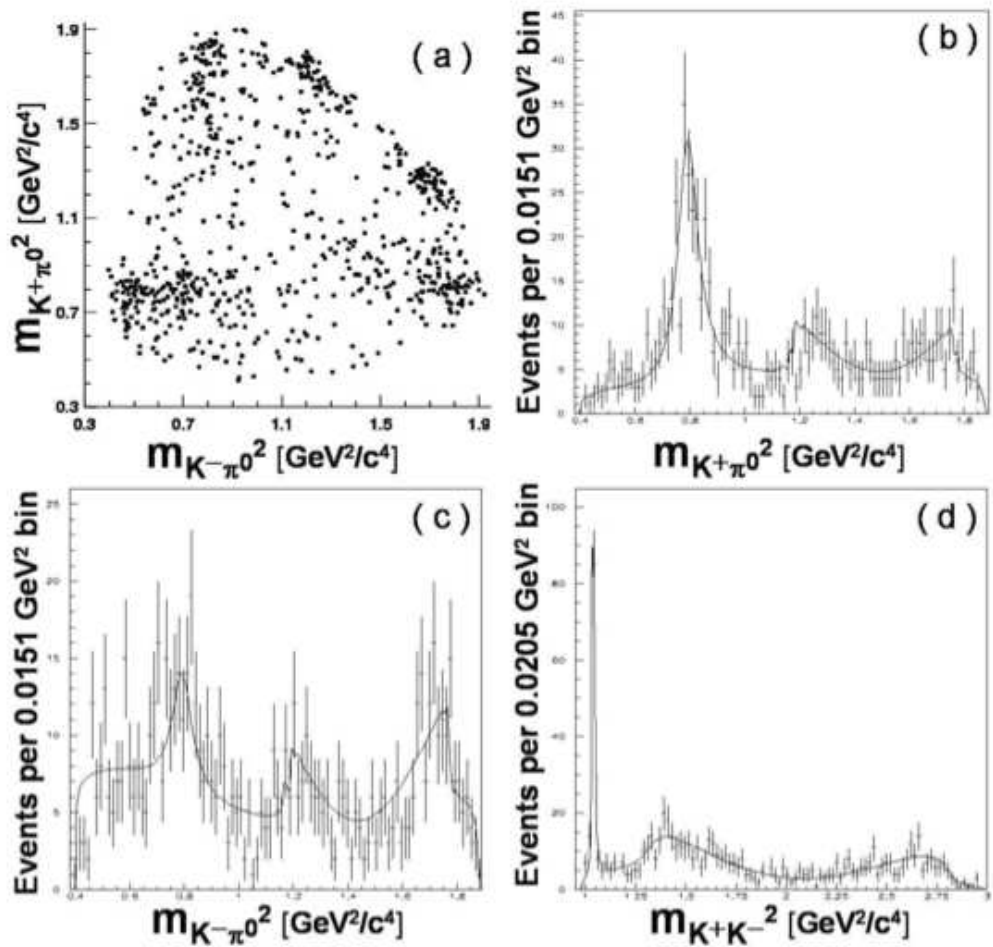}
\caption{Dalitz plot and projections for $D^0 \to K^+K^-\pi^0$} 
\label{fig:kkpi0dalitz}
\end{figure*}

	The analysis is done with CLEO-III data and the $D^0$'s are produced
in continuum production with their flavor tagged by the charge of the soft
pion in charged $D^\ast$ decays.  The sample is about 600 events with a signal
to noise of about two to one.  For the Dalitz plot analysis the charges
are swapped such that all the decays are treated as if a $D^0$ is decaying.
The Dalitz plot as shown in Figure~\ref{fig:kkpi0dalitz}
shows clear contributions from both charges of $K^\ast$ plus an oppositely
charged $K$ and $\phi-\pi^0$.
The contributions to the Dalitz plot are adequately described by just
these three contributions plus a non-resonant amplitude which is
assumed to be uniform in phase space and have a fixed phase which does
not interfere with the resonant amplitudes.  The projections of the Dalitz
plot on the three mass combinations are shown in Figure~\ref{fig:kkpi0dalitz}
along with the result of a fit with the contributions described above.

	Many systematic effects are considered.  The largest effects are
caused by considering other resonance contributions to the Dalitz plot.
None are found to be significant, but they do change the central values
for the clearly observed resonances.  The background is modeled with mass
and mass difference side bands that are then fit to second order two-dimensional
polynomials.  Allowing the parameters describing the background to vary is the
next largest systematic
uncertainty.  The efficiency across the Dalitz plot is modeled with a simulated
sample that is generated uniformly across the Dalitz plot and those passing
all the analysis selections are fit to a two dimensional second order polynomial.
Again allowing these parameters to vary provide the third largest contribution
to the systematic uncertainty.  
The effect of varying analysis details and procedures on the systematic
uncertainty is small compared to the three effects described above.

	The analysis is preliminary.  Table~\ref{tab:kkpi0dalitz} gives the
\begin{table*}[ht]
\begin{center}
\caption{Results of the Dalitz plot analysis of $D^0 \to K^+K^-\pi^0$.  The
         errors are only statistical.}
\begin{tabular}{|l|c|c|c|}
\hline 
\textbf{Contribution} & \textbf{Amplitude} & \textbf{Phase (${}^\circ$)} & \textbf{Fit Fraction (\%)} \\ \hline
$K^{\ast +}K^-$       & $1.0$ (fixed)      & $0.0$ (fixed)               & $46.1\pm3.1$ \\
$K^{\ast -}K^+$       & $0.52\pm0.05$      & $332\pm8$                   & $12.3\pm2.2$ \\
$\phi\pi^0$           & $0.64\pm0.04$      & $326\pm9$                   & $14.9\pm1.6$ \\
Non-resonant          & $5.62\pm0.45$      & $220\pm5$                   & $36.0\pm3.7$ \\ \hline
\end{tabular}
\label{tab:kkpi0dalitz}
\end{center}
\end{table*}
amplitude and phase relative to the $D^0 \to K^{\ast +}K^-$ contribution
for the other contributions, and the fit fraction for all the contributions.
Only statistical errors are shown.  The phase between $D^0 \to K^{\ast +}K^-$
and $D^0 \to K^{\ast -}K^+$ is measured as $(332 \pm 8 \pm 11)^\circ$
and the amplitude for $K^{\ast -}K^+$ relative to $K^{\ast +}K^-$ is 
$0.52 \pm 0.05 \pm 0.04$.  Even with this small data set the precision
is limited by non-$K^\ast K$ contributions to the decay.

\section{Dalitz Plot Analysis of $D^+ \to \pi^+\pi^-\pi^+$}

	A Dalitz plot analysis of $D^+ \to \pi^+\pi^-\pi^+$ has previously been done by
E791~\cite{791} and~FOCUS~\cite{focus}.  The analysis reported on here is from
CLEO-c, and represents the first time CLEO has done the same Dalitz plot analysis
as the fixed target experiments.  Previously CLEO has focused on analyses with
$\pi^0$'s in the final state.  The decay is selected with cuts on the beam constrained
mass of three charged tracks consistent with pions and the difference of their energy
from the beam energy.  A sample of 2600 events is selected with a signal to noise
of better than two to one.  The E791 and FOCUS samples are of similar size and cleanliness.

	The Dalitz plot is symmetric under the interchange of like-sign pions thus the
analysis is done in the two dimensions of high unlike-sign pion mass squared versus
low unlike-sign pion mass.  There is a large contribution from $K_{\rm S}\pi$ which
because of the long $K_{\rm S}$ lifetime should not interfere with the other contributions
to the plot.  This stripe on the Dalitz plot is not considered when a fit is done
for two body resonance contributions.
The efficiency across the Dalitz plot is modeled with simulated events that is
then fit to a two-dimensional second order polynomial.  While there is a notable
fall of the efficiency in the corners of the Dalitz plot the changes are smooth,
and well modeled by the polynomial.  Backgrounds are taken from sidebands and 
extra resonance contributions are allowed from mismeasured $K_{\rm S}$, $\rho$,
and $f^0(1370)$ decays.  Many possible resonances can contribute to the decay,
and a total of 13 different resonances are considered.  Parameters describing these
resonances are taken from previous experiments. Only those with an
amplitude significant at more than three standard deviations are said to
be observed, and others are limited.

\begin{figure*}[t]
\includegraphics[width=55mm]{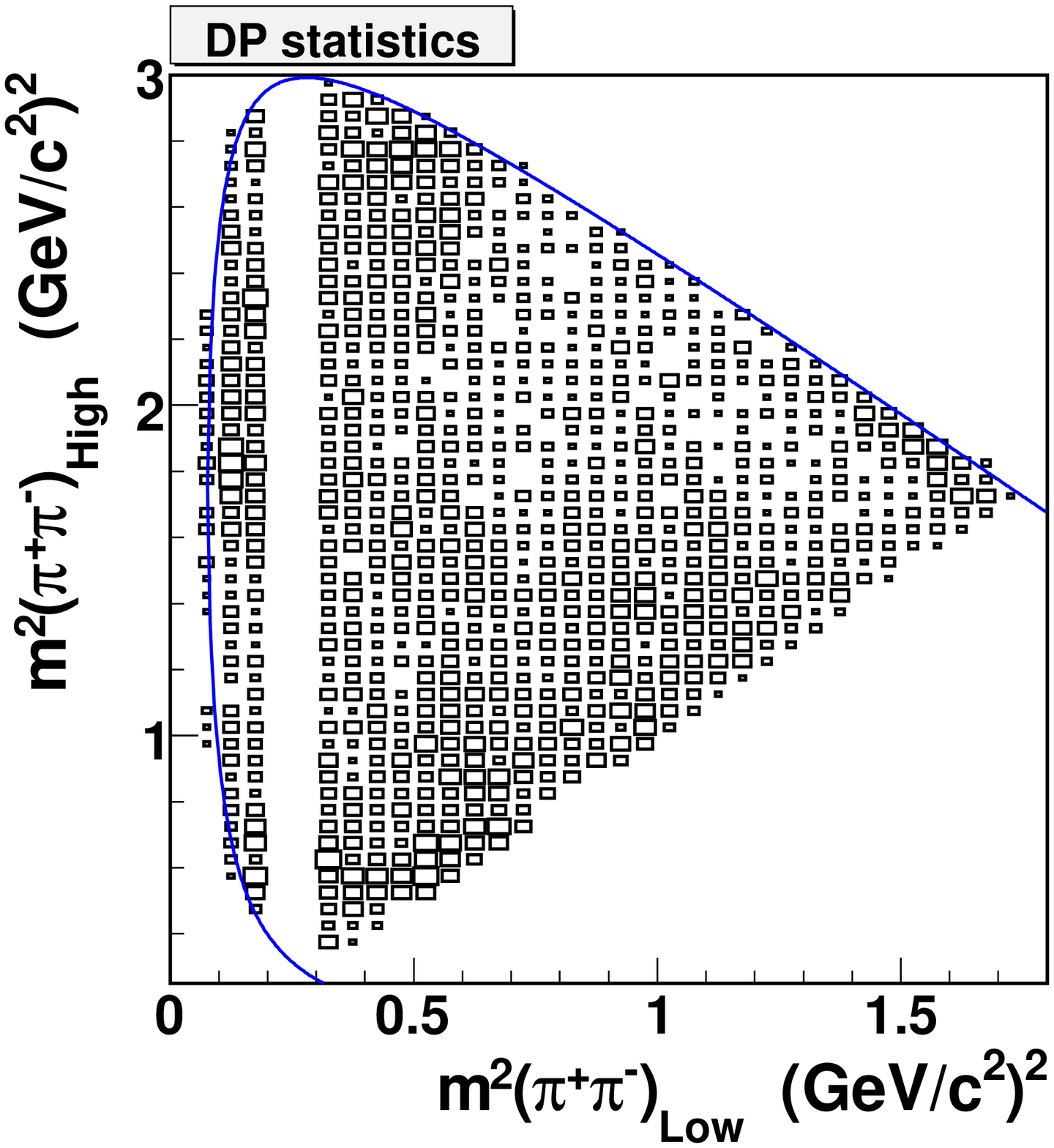}
\includegraphics[width=55mm]{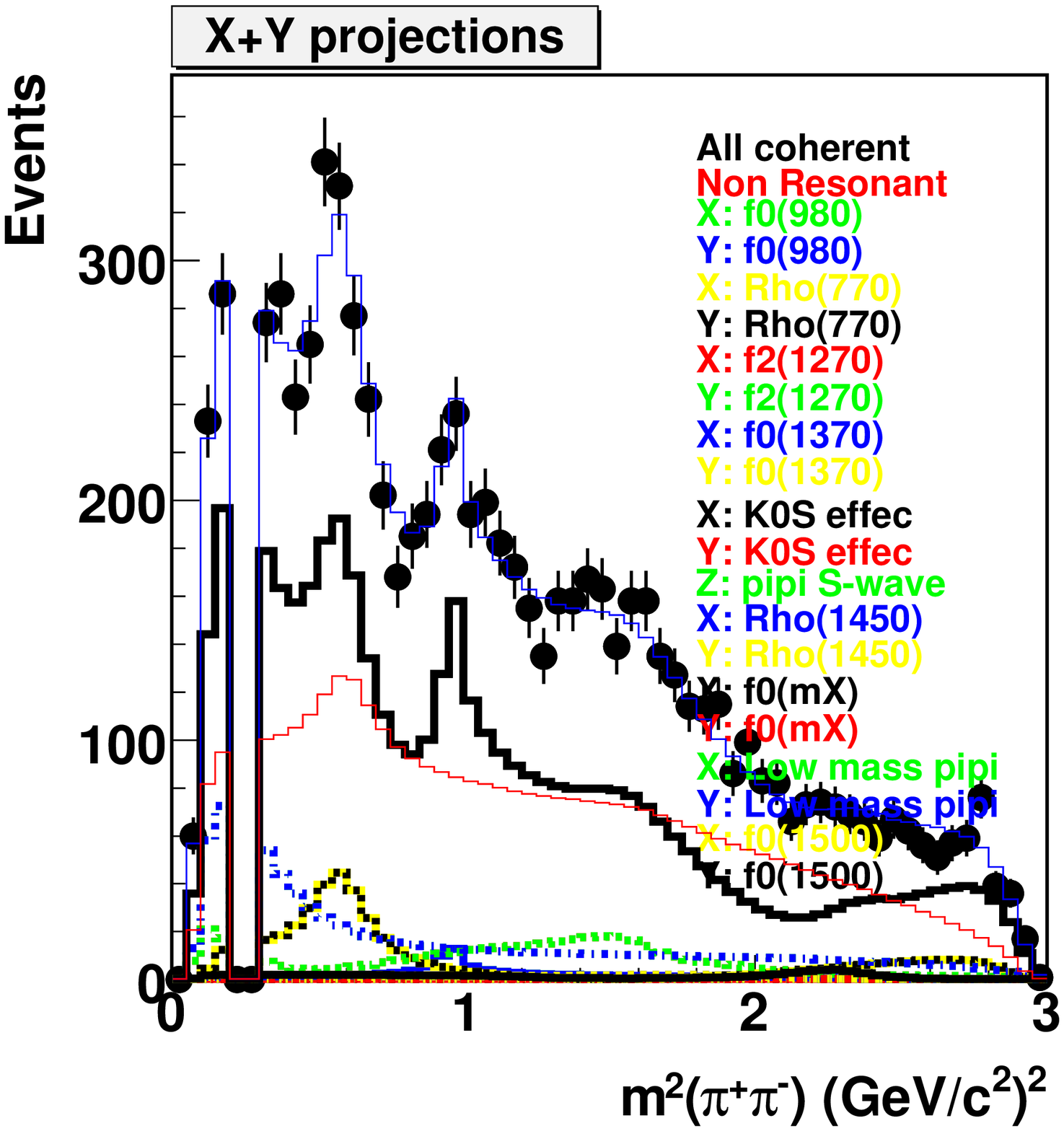}
\includegraphics[width=55mm]{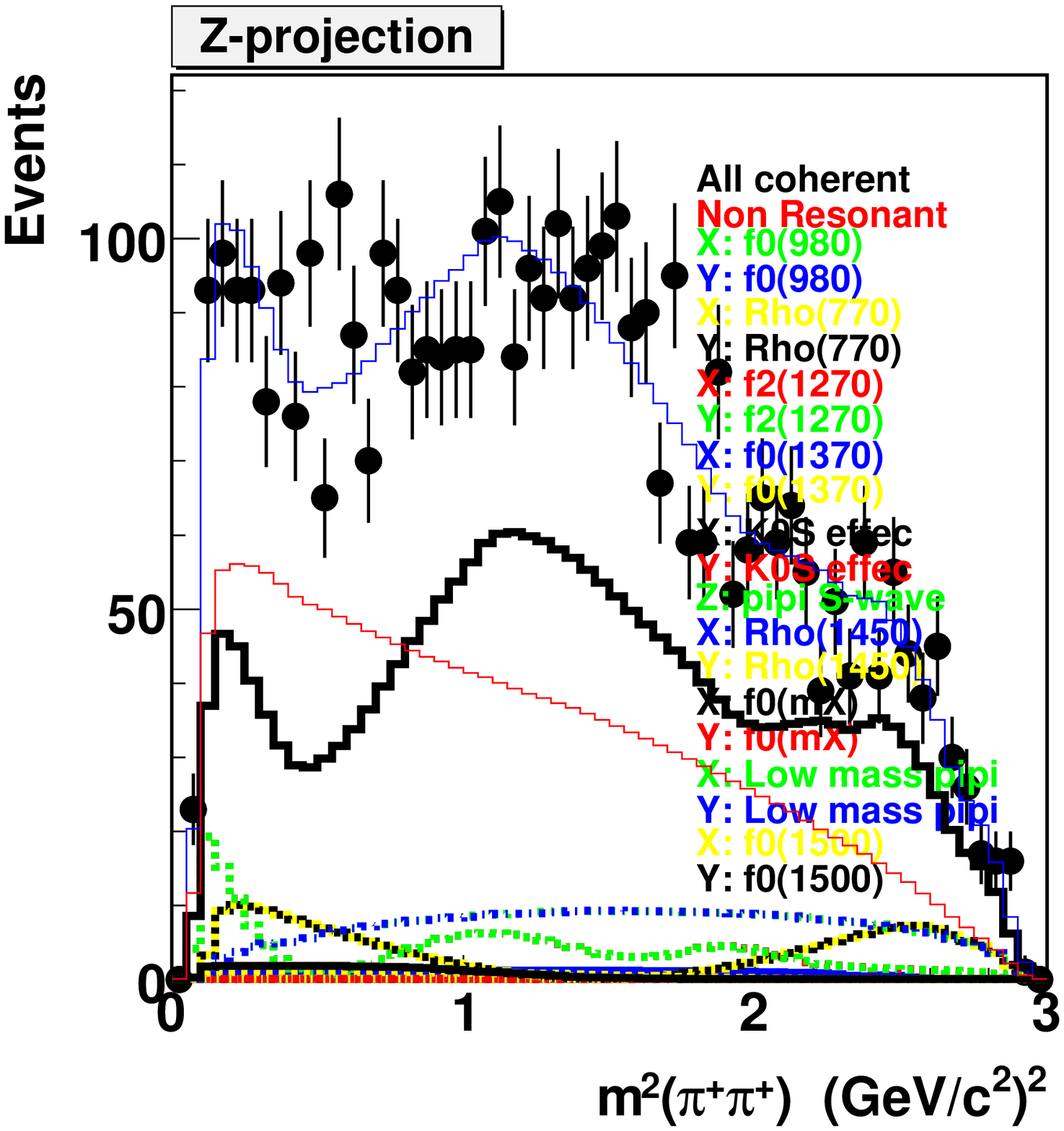}
\caption{Dalitz plot and projections for $D^+ \to \pi^+\pi^-\pi^+$} 
\label{fig:d3pidalitz}
\end{figure*}

	Figure~\ref{fig:d3pidalitz} shows the Dalitz plot and projections on to
the squared masses.  Contributions from $\rho^0\pi$ and $f_2(1270)\pi$ are clearly
visible.  Table~\ref{tab:d3pidalitz} shows the fit fractions measured by CLEO
\begin{table}[t]
\begin{center}
\caption{Comparison of the fit fractions, in percent, found in the Dalitz plot analysis of $D^+ \to \pi^+\pi^-\pi^+$ between
         CLEO and E791.  CLEO limits are at 90\% confidence level.}
\begin{tabular}{|l|c|c|}
\hline 
\textbf{Contribution} & \textbf{CLEO} & \textbf{E791} \\ \hline
$\rho^0 \pi^+$        & $20.0\pm2.5$  & $33.6\pm3.9$ \\
$\sigma^0 \pi^+$      & $41.8\pm2.9$  & $46.3\pm9.2$ \\
$f_2(1270) \pi^+$     & $18.2\pm2.7$  & $19.4\pm2.5$ \\
$f_0(980) \pi^+$      &  $4.1\pm0.9$  & $6.2\pm1.4$ \\
$f_0(1500) \pi^+$     &  $3.4\pm1.3$  & ${}$ \\
Non-resonant          &  $< 3.5$      & $7.8\pm6.6$ \\
$\rho(1450) \pi^+$    &  $< 2.4$      & $0.7\pm0.8$ \\ \hline
\end{tabular}
\label{tab:d3pidalitz}
\end{center}
\end{table}
comparing with the results of the E791 analysis mentioned above.  There is broad
agreement between the two results, including the observation of a $\sigma\pi$
contribution.  The CLEO analysis is preliminary, and they plan to consider a generalized
model of $\pi\pi$ S-wave interactions to model $\sigma$ and $f_0$ contributions
such as the K-matrix which is used in the FOCUS analysis mentioned above.

\section{Quantum Correlations in $\psi(3770) \to D^0 \bar D^0$}

	$D^0$ meson pairs produced in $e^+e^- \to \psi(3770)$ are in a C~$-$ eigenstate.
Thus if both $D$'s decay to a mode that tags of the flavor of the meson, such as $D^0 \to K^-\pi^+$
versus $\bar {D^0} \to K^+\pi^-$ there will be interference.  The total rate for such double 
tags will depend on the DCS rate and the strong phase between the Cabibbo favored and DCS
decay.  Double flavor tags with the same mode and flavor are forbidden unless there is $D$ mixing,
and thus the rate for these depends on the mixing parameters $x$, related to the width difference,
and $y$, related to the mass difference.  Flavor tags opposite semileptonic decays, which are not
eigenstates of C, measure the isolated decay rates without the interference effects and have
different sensitivity to mixing and the DCS rate.  $D^0$ decays to CP eigenstates opposite
a flavor tag have yet a different dependence on mixing and the DCS rate.  Double CP tags show
maximal interference effects with like CP's being forbidden and opposite CP's being enhanced.
The expected rates are summarized in Table~\ref{tab:tqca1} and explained in full detail in~\cite{AsnerandSun}.
\begin{table}[h]
\begin{center}
\caption{The effect of quantum correlations in $\psi(3770)$ decays to $D^0$ pairs.  The columns and rows
         represent the decay modes of one $D$ with $f$ representing a hadronic mode that tags the flavor, $\ell$ a
         semileptonic decay, CP a mode that is an eigenstate of CP, and X is any mode.  $x$ and $y$ are the
         $D$ mixing parameters, $R_M = (x^2 + y^2)/2$, $r = $ Amplitude for DCS/Amplitude for Cabibbo Favored,
         and $\delta$ is the strong phase between the Cabibbo favored and DCS decay.}
\begin{tabular}{|c|cccc|}
\hline 
         & $f$                             & $\ell+$ & CP$+$ & CP$-$ \\ \hline
$f$      & $R_M/r^2$                       &         &       &       \\
$\bar f$ & $1 + r^2(2 - (2\cos\delta)^2)$  &         &       &       \\
$\ell-$  & $1$                             & $1$     &       &       \\
CP$+$    & $1 + r (2\cos\delta)$           & $1$     & $0$   &       \\
CP$-$    & $1 - r (2\cos\delta)$           & $1$     & $2$   & $0$   \\ \hline
X        & $1 + r y (2\cos\delta)$         & $1$     & 1 - y & 1 + y \\ \hline
\end{tabular}
\label{tab:tqca1}
\end{center}
\end{table}
Simultaneously this analysis can measure branching fractions which are presently
poorly known for many CP eigenstates.  

	A preliminary version of this analysis is done by CLEO-c.
The analysis has three major pieces.  Single flavor tags and CP tags are identified using
the methods developed for precision measurement of $D$ branching fractions~\cite{Dbranch},
and similarly for hadronic double tags.  Semileptonics are only measured opposite
flavor and CP tags and are generally very clean in flavor tags, but more difficult in
CP tags.  The effects of the quantum correlation are clearly seen in double CP tags where
for three CP$+$ modes ($K^+K^-$, $\pi^+\pi^-$, and $K_{\rm S}\pi^0\pi^0$) and one
CP$-$ mode ($K_{\rm S}\pi^0$) $29.6\pm0.9$ like CP tags would be expected in the
absence of any correlations and $5.3\pm3.7$ are observed.  For opposite CP
tags $29.1\pm0.8$ are expected in the absence of quantum correlation and $72.6\pm8.5$
are observed.  Table~\ref{tab:tqca2} shows results with statistical errors
\begin{table}[h]
\begin{center}
\caption{Preliminary results from the CLEO-c quantum correlation analysis outlined in the text.
         The errors from CLEO are only statistical.}
\begin{tabular}{|l|cc|} \hline 
Parameter         & CLEO QCA         & Present Knowledge \\ \hline
$y$               & $-0.057\pm0.066$ & $0.008\pm0.005$ \\
$r^2$             & $-0.028\pm0.069$ & $0.00374 \pm 0.00018$ \\
$r (2\cos\delta)$ & $0.130\pm 0.082$ &                       \\
$R_M$             & $(1.74\pm1.47) \times 10^{-3}$ & $<\sim 1 \times 10^{-3}$ \\ \hline
\end{tabular}
\label{tab:tqca2}
\end{center}
\end{table}
for mixing and DCS parameters and the strong phase from the full analysis compared
with the present knowledge from other experiments.  While at this stage the analysis is
not competitive CLEO projects that with their full data set it will be producing
results with a precision similar to world averages and with an orthogonal set of systematic
uncertainties. This analysis depends on the assumption that the $D$ pairs observed
from the $\psi(3770)$ are in the C~$-$ state, and they preliminarily measure the rate
of CP$+$ states from $\psi(3770)$ decay at $0.06\pm0.05$, where the error is 
only statistical.  Theoretical expectations
are orders of magnitude below this.

\section{Conclusion}

	Quantum correlations in charm decays provide a unique opportunity
to study the details of decay dynamics.  Here I have shown two recent Dalitz
analyses that build on previous work and are probing strong phases and strong dynamics at new levels
of precision.  Also I have shown how quantum correlations in threshold production
give unique leverage and provide alternative ways to measure rare behavior
in the charm sector.  We can hope that results in this fascinating field
continue to multiply to enlighten and challenge our colleagues.

\section{Acknowledgment}

	Thanks to the organizers for a very well run, interesting, and enjoyable conference.
Specific thanks to Jeff Appel, David Asner, and Chris Hearty.  My research activities
are supported by the U.S. National Science Foundation.

\end{document}